\documentclass[utf8]{frontiersSCNS} 

\usepackage{url,hyperref,lineno,microtype,subcaption}
\usepackage[onehalfspacing]{setspace}
\usepackage{multirow}
\usepackage{rotating}
\usepackage{tikz}

\usepackage{xcolor}

\def\keyFont{\fontsize{8}{11}\helveticabold }
\def\firstAuthorLast{M. B. Kors\'os {et~al.}} 

\def\Authors{M. B. Kors\'os\,$^{1,2,3}$, R. Erd{\'e}lyi\,$^{4,2,3}$,  J. Liu\,$^{5}$ and H. Morgan\,$^{1}$ }

\def\Address{$^{1}$Department of Physics, Aberystwyth University, Ceredigion, Cymru, SY23 3BZ, UK\\
$^{2}$Department of Astronomy, E\"otv\"os Lor\'and University, P\'azm\'any P\'eter s\'et\'any 1/A, H-1112 Budapest, Hungary\\
$^{3}$Gyula Bay Zolt\'an Solar Observatory (GSO), Hungarian Solar Physics Foundation (HSPF), Pet\H{o}fi t\'er 3., Gyula, H-5700, Hungary\\
$^{4}$Solar Physics \& Space Plasma Research Center (SP2RC), School of Mathematics and Statistics, University of Sheffield, Hounsfield Road, S3 7RH, UK\\
$^{5}$ Astrophysics Research Centre (ARC), School of Mathematics and Physics, Queen’s University, Belfast, BT7 1NN, NI, UK}

\def\corrAuthor{R. Erd{\'e}lyi}
\def\corrEmail{robertus@sheffield.ac.uk}
\begin{document}
\onecolumn
\firstpage{1}

\title[Flare prediction by the $S_{l-f}$ and $G_{S}$ proxies ]{Testing and Validating Two Morphological Flare Predictors by Logistic Regression Machine Learning}

\author[\firstAuthorLast ]{\Authors} 
\address{} 
\correspondance{} 

\extraAuth{}
\maketitle

\begin{abstract}
 Whilst the most dynamic solar active regions (ARs) are known to flare frequently, predicting the occurrence of individual flares and their magnitude, is very much a developing field with strong potentials for machine learning applications.

The present work is based on a method which is developed to define numerical measures of the mixed states of ARs with opposite polarities. The method yields compelling evidence for the assumed connection between the level of mixed states of a given AR and the level of the solar eruptive probability of this AR by employing two morphological parameters: (i) the separation parameter $S_{l-f}$ and (ii) the sum of the horizontal magnetic gradient $G_{S}$.

In this work, we study the efficiency of $S_{l-f}$ and $G_{S}$ as flare predictors on a representative sample of ARs, based on the SOHO/MDI-Debrecen Data (SDD) and the SDO/HMI - Debrecen Data (HMIDD) sunspot catalogues.
In particular, we investigate about 1000 ARs in order to test and validate the joint prediction capabilities of the two morphological parameters by applying the logistic regression machine learning method.
Here, we confirm that the two parameters with their threshold values are, when applied together, good complementary predictors. Furthermore, the prediction probability of these predictor parameters is given at least 70\% a day before. 
\tiny
 \keyFont{ \section{Keywords:} flare prediction- morphological parameters -validation-  binary logistic regression- machine learning} 
 \end{abstract}

\section{Introduction}

A solar flare is a sudden flash observed in the solar atmosphere which is able to rapidly heat the plasma to megakelvin temperatures, while the electrons, protons and other heavier ions are accelerated to very large speeds \citep{Benz2008}. The associated accelerated particle clouds may reach the Earth, typically within a few hours or a day following a solar flare eruption. The flares produce radiation across the electromagnetic spectrum at all wavelengths. Most of the released energy is spread over frequencies outside the visible range. For this reason, the majority of flares must be observed with instruments which measurements in these wavelength ranges, as e.g. the Geostationary Operational Environmental Satellite (GOES). Therefore, the most generally known flare classification scheme is GOES flare-class. Measurements of the maximum x-ray flux at wavelengths from 0.1 to 0.8 nm near Earth are classed as A, B, C, M, or X type flares back from 1975\footnote{http://hesperia.gsfc.nasa.gov/goes/goes\_event\_listings/}. These five GOES flare intensity categories are further divided into a logarithmic scale labelled from 1 to 9. The A-, B- and C-classes are the lowest energy release classes of solar flares and they also occur frequently in the solar atmosphere. The A to C-class range has no or hardly any detectable effect on Earth based on current instrumentations and understanding. The M-class medium flare category may cause smaller or occasionally more serious disruptions, e.g. radio blackouts. However, the X-intensity flares may cause strong to extreme hazardous events, facility break-downs (e.g. radio blackouts, etc.) on the daylight side of the Earth \citep{Hayes2017}. The major solar flares (M- and X-class) are often accompany with accelerated solar energetic particles and coronal mass ejections (CMEs) \citep[see, e.g., ][]{Tziotzio2010}.

For solar activity modelling, a key ingredient is to determine the role of the associated observable magnetic field. \cite{Waldmeier1938} proposed the first classification scheme to examine the connection between the size and morphology of active regions (ARs) and the capacity of their flare-productivity. This classification scheme is known today as the Z\"urich classification \citep[see also][]{Kiepenheuer1953}. This scheme contains eight types thought to be representative of consecutive states in the evolution of a sunspot group. The classification system was further developed by \cite{McIntosh1990}. McIntosh introduced three more components based on characteristics including the Z\"urich class, the largest sunspot, and the sunspot distribution in an AR. Although the classification uses white-light observations only, it is still widely used.

The first magnetic classification scheme, known as the Mount Wilson classification, was introduced by \cite{Hale1919}. It is simpler than the Z\"urich-McIntosh system, as it only distinguishes unipolar, bipolar, mixed configurations and very close and mixed configurations within a common penumbral feature, denoted by the letters $\alpha$, $\beta$, $\gamma$ and $\delta$-class, respectively. \cite{Kunzel1960} added the $\delta$-class configurations for the McIntosh system which refer to the most productive sources of energetic flares \citep[see, e.g., ][and references therein]{Schrijver2016}. 
 All these classification schemes are useful in revealing potential connections between the morphological  properties of sunspot groups and their flare-productivity. However, it is somewhat ambiguous that these classification schemes rely on a number of rather subjective elements to be identified by visual inspection besides some more objective measures.

The McIntosh and Mount Wilson classifications have been shown to be useful for grouping ARs by their expected flare productivity \citep{Gallagher2002,Ireland2008,Bloomfield2012}. 
However, further quantities derived from AR observations allow a physical comparison and deeper understanding of the actual causes of the solar eruptions. In this sense, different morphological parameters have been introduced to characterised the magnetic field configuration or highlight the existence of polarity-inversion-lines (PILs) in ARs, with varying sophistication \citep[see e.g.][and references therein]{Barnes2016, Leka2018, Leka2019a,Leka2019b,Campi2019, Park2020}.
Furthermore, \cite{Kontogiannis2018} investigated and tested some of those parameters, which were identified as efficient flare predictors. These parameters include, e.g., a quantity denoted as $B_{eff}$ that measures the coronal magnetic connectivity between the opposite magnetic field elements \citep{Georgoulis2007}, Ising energy $E_{Ising}$ of a distribution of interacting magnetic elements \citep{Ahmed2010}, the sum of the horizontal magnetic field gradient $G_{S}$ \citep{Korsos2016}, and the total unsigned non-neutralized currents, $I_{NN,tot}$ \citep{Kontogiannis2017}. 

The observed magnetic properties of an AR can be processed for the purpose of prediction by machine learning (ML) computational methods for data analysis \citep{Camporeale2019}, such as neural networks \citep{Ahmed2013}, support vector machines \citep{Bobra2015,Boucheron2015}, relevance vector machines \citep{Al-Ghraibah2015}, ordinal logistic regression \citep{Song2009}, decision trees \citep{Yu2009}, random forests \citep{Liu2017,Domija2019}, and deep learning \citep{Nishizuka2018}. Notably, parameters $B_{eff}$, $E_{Ising}$, $G_{S}$, and $I_{NN,tot}$ were used by the FLARECAST project\footnote{http://flarecast.eu}, where the prediction capabilities of almost 200 parameters were tested by the LASSO and Random Forest ML techniques \citep{Campi2019}. From these 200 parameters, the FLARECAST project found that the four morphological parameters were ranked as good flare predictors. 

The content of the paper is as follows: Section~\ref{method} overviews in detail the two morphological parameters used for flare prediction in this work. Section~\ref{data} describes the data preparation process and key aspects of the adopted ML method. Section \ref{analyses} shows the results of the analysis focusing on two morphological parameters in particular, while our conclusions are in Section \ref{conclusion}.

 \section{Two morphological parameters} \label{method}
  
 \begin{figure}
\centering
\includegraphics[width=1\textwidth]{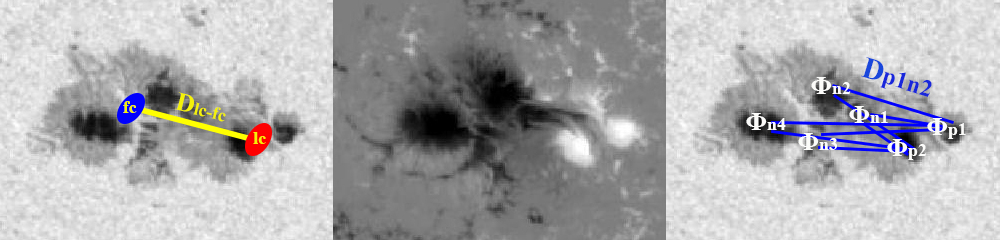}
\put(-500,105){(a)}
\put(-330,105){\textcolor{white}{(b)}}
\put(-165,105){(c)}

\caption{\label{Fig} Figures illustrating the determination of the  $S_{l-f}$ and $G_{S}$ morphological parameters. Panel (a) demonstrates, for $S_{l-f}$, how the distance $D_{lc-fc}$ is taken between the area-weighted centres (therefore the index {\it c}) of the spots of leading $l$ and following $f$ polarities. Panel (b) is the corresponding magnetogram of the continuum image of AR 11775, which were taken at 00:59 on 20 June 2013. Panel (c) present how the $G_{S}$ parameter is calculated. $\Phi$ is the magnetic flux in a positive $p$ or negative $n$ umbra. $D$ is the distance between two opposite-polarity umbrae.}
\end{figure}

\cite{Korsos2016} introduced and tested, as a trial, an advantageous scheme that may be used as new prediction indicators besides the  Z\"urich, McIntosh and  Mount Wilson classification systems. This scheme includes two morphological parameters, namely: 

\begin{itemize}

\item The separation parameter $S_{l-f}$, which characterises the separation of opposite polarity subgroups in an AR, given by the formula:
 
  \begin{equation}
    S_{l-f}=\frac{D_{lc-fc}}{2\sqrt{{\sum A_{g}/\pi}}},
    \label{saparatness}
 \end{equation}
 \noindent
where $l$ and $f$ refer to the leading and following polarities. The numerator denotes the distance between the area-weighted centres (therefore the index {\it c}) of the spots of leading and following polarities. Fig.~\ref{Fig}a gives a visual representation. The denominator is the diameter of a hypothetic circle (2 times the radius ($ \sqrt{{\sum A_{g}/\pi}}$)). The $\sum$$A_{g}$ is the sum of individual umbrae areas in a sunspot group.

\item The second introduced morphological parameter is the sum of the horizontal magnetic gradient $G_{S}$, defined by
  \begin{equation}
	G_{S}=\sum_{i,j} \frac{\left |\Phi_{p,i}-\Phi_{n,j}\right |}{D_{i,j}},
	\label{sumgrad}
 \end{equation}
 \noindent
where $\Phi$ is the magnetic flux of the umbra based on \cite{Korsos2014}. The indices $p$ and $n$ denote positive and negative polarities, and $i$ and $j$ are their running indices in the entire sunspot group. $D$ is the distance between two opposite-polarity umbrae with indices $i$ and $j$, respectively. Panel c of Fig.\ref{Fig} gives a visual presentation of the $G_{S}$ parameter.

 \end{itemize}

 The $S_{l-f}$ and $G_{S}$ can be determined from the moment of first available observation of sunspot groups, because the applied umbrae data are suitably corrected for geometrical foreshortening in the SOHO/MDI-Debrecen Data (SDD\footnote{http://fenyi.solarobs.csfk.mta.hu/en/databases/SOHO/}) and the SDO/HMI - Debrecen Data (HMIDD\footnote{http://fenyi.solarobs.csfk.mta.hu/en/databases/SDO/}) catalogues \citep{Baranyi2016}. Furthermore, these two morphological parameters were shown to be potential indicators for upcoming flares on a smaller number of typical test cases \citep{Korsos2016}. The test cases included 116 ARs, which were selected from SDD. Their selection was based on that about a third of the ARs produced only B- and C-class flares, another third produced M-class flares, and the remaining third produced X-class flares. For the statistical analysis, the considered values of $S_{l-f}$ and $G_{S}$ were determined 24, 48, and 72 hrs before flare onset to test the conditional flare probability (CFP) of these two parameters. The CFPs were calculated as empirical probabilities, which measure the studied flare intensities and adequate recordings of the happening of events. 
 
\cite{Korsos2016} found that if $S_{l-f}$$\leq$1 for a flaring AR then the CFP of the expected largest intensity  flare being X-class is over at least 70\%. If 1$\leq$$S_{l-f}$$\leq$3 the CFP is more than 45\% for the largest-intensity flare(s) to be the M-class, and, if 3$\leq$$S_{l-f}$$\leq$13 there is larger than 60\% CFP that C-class flare(s) may occurs within a 48-hr interval. 
Next, \cite{Korsos2016} found also that from analysing $G_{S}$ independently for determining the associated CFPs:  if 7.5$\leq$$\log(G_{S})$ then there is at least 70\% chance for the strongest energy release to be X-class; if 6.5$\leq$$\log(G_{S})$$\leq$7.5 then there is $\sim$45\% CFP that M-class could be the highest-intensity flares; finally, if 5.5$\leq$$\log(G_{S})$$\leq$6.5, then it is very likely that C-class flare(s) may be the main intensity flares in the coming 48 hours. ARs are unlikely to produce X-class flare(s) if 13$\leq$$S_{l-f}$ and $\log$($G_{S}$)$\leq$5.5.

\section{Data and data preparation} \label{data}

\begin{figure}
\centering
\includegraphics[width=0.49\textwidth]{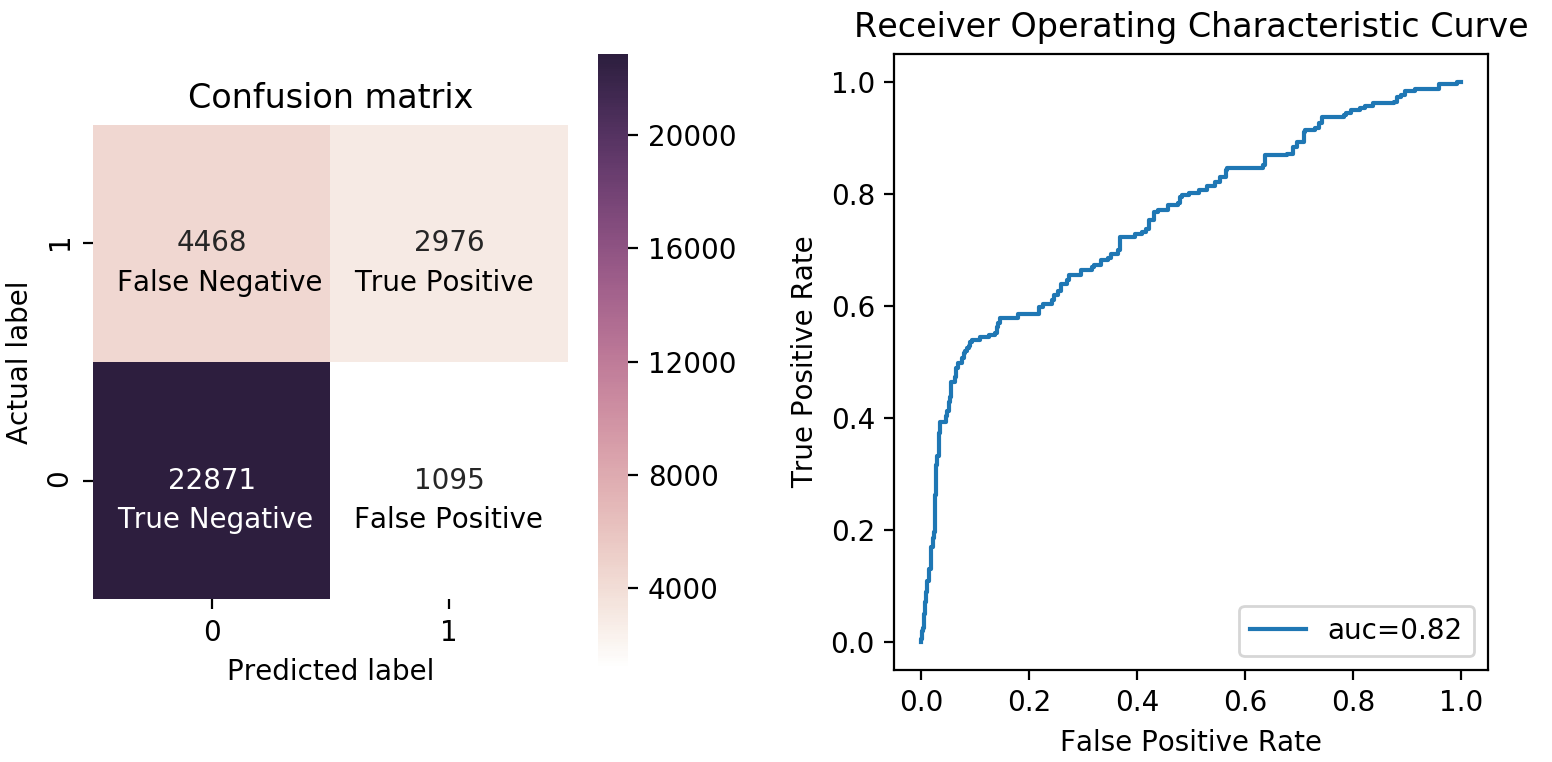}
\put(-240,110){(a)}
\includegraphics[width=0.49\textwidth]{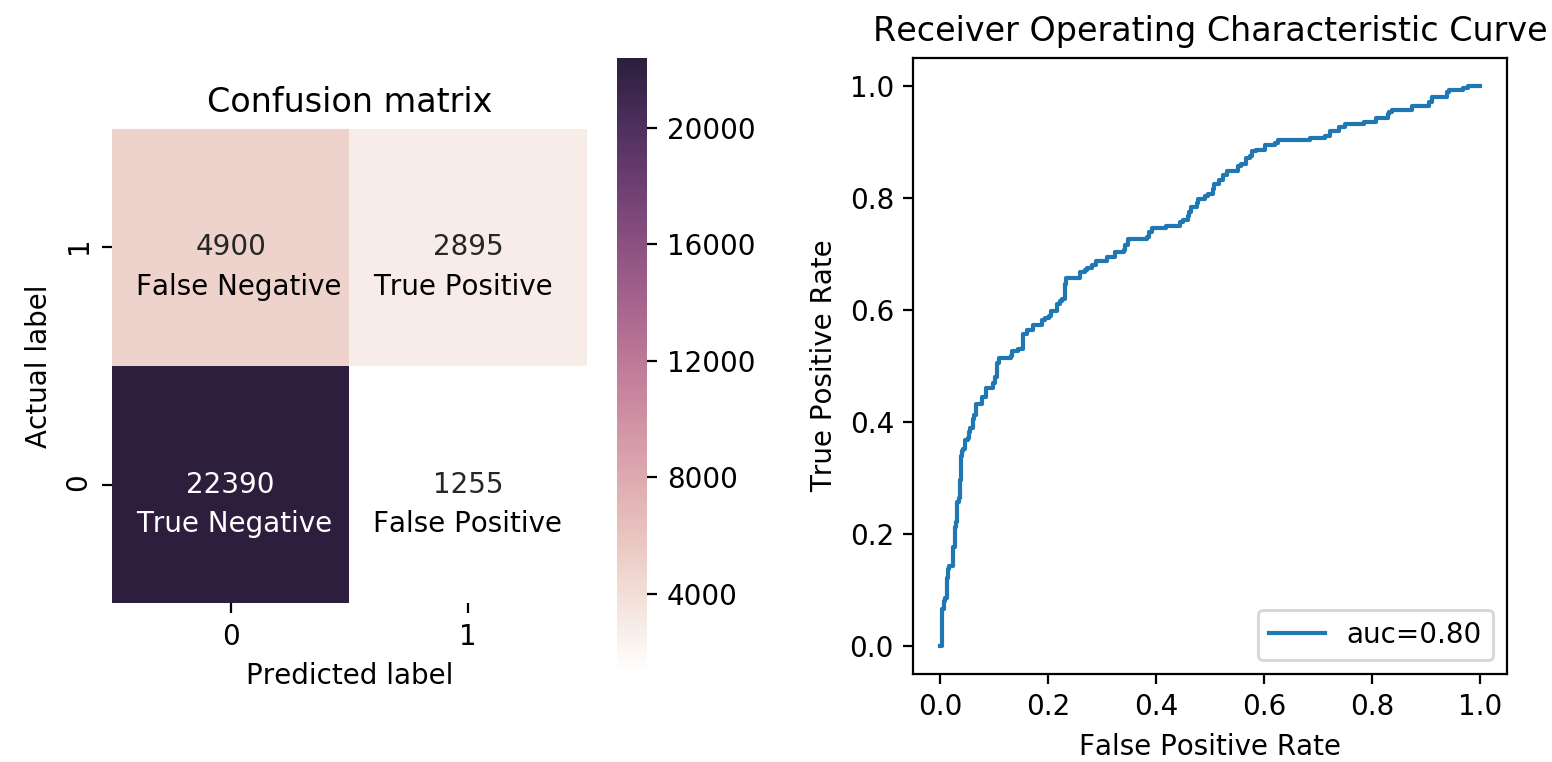}
\put(-240,110){(b)}\\
\includegraphics[width=0.49\textwidth]{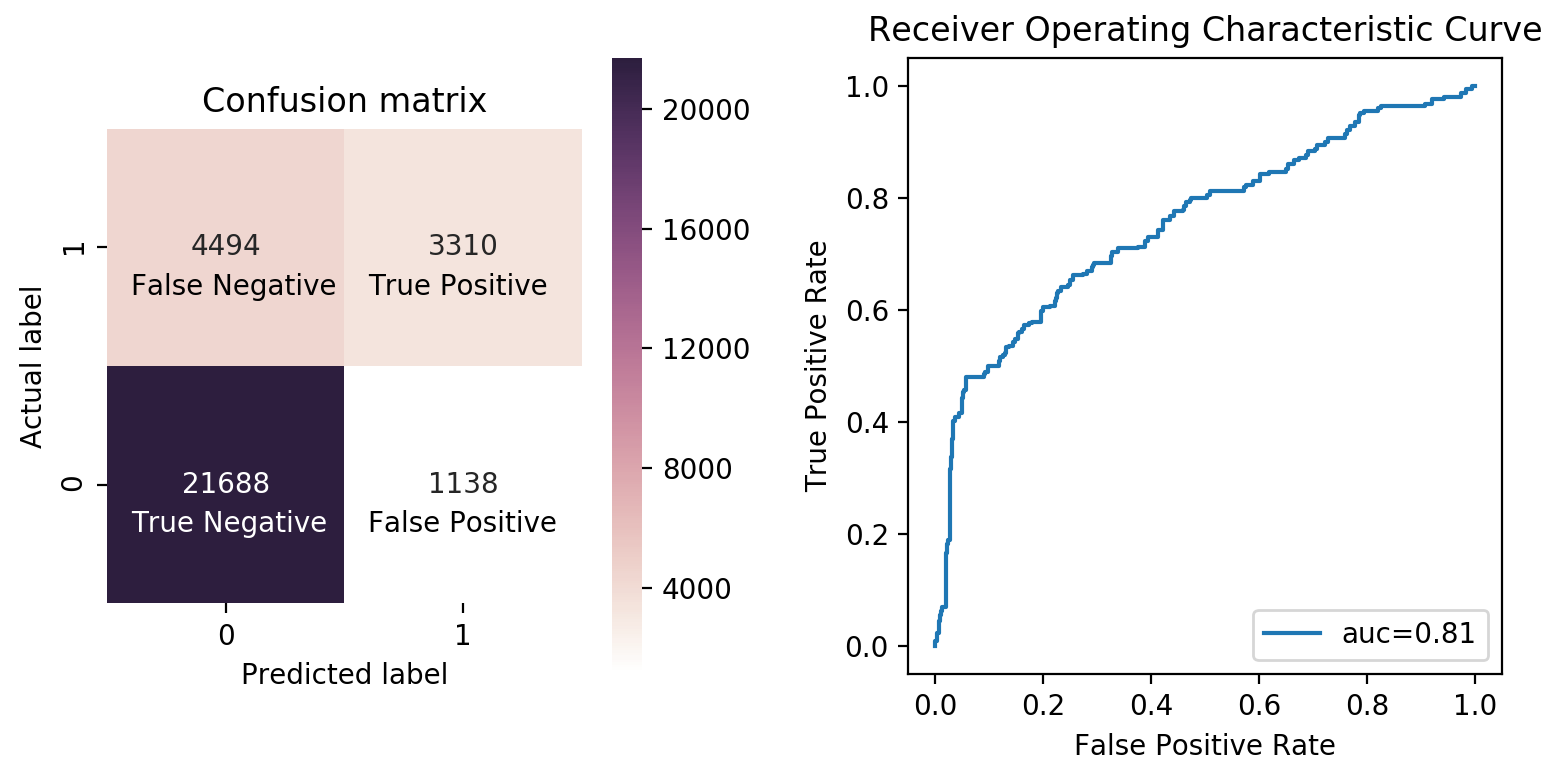}
\put(-240,110){(c)}
\includegraphics[width=0.49\textwidth]{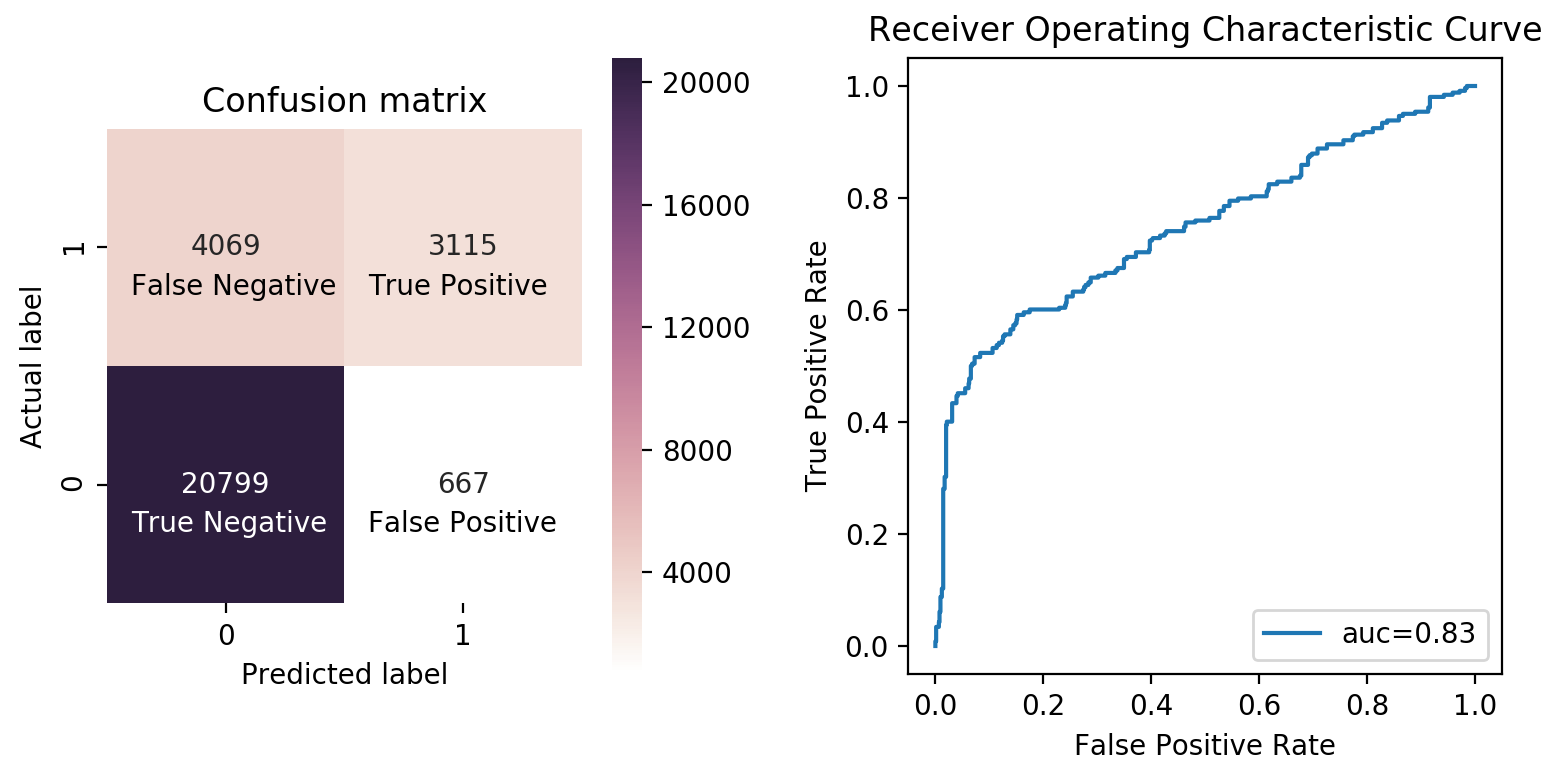}
\put(-240,110){(d)}
\caption{\label{Flare} The result of the binary logistic regression of the 1$^{st}$ model with 6-, 12-, 18-, and 24-hr forecast issuing times for panels (a), (b), (c), and (d) respectively. The right side of each panel presents the corresponding Receiver Operating Characteristic (ROC) curves.}
\end{figure}

In this study, we further explore test and validate, the joint prediction capabilities of the $S_{l-f}$ and $G_{S}$ morphological parameters. The analysis is based on the binary logistic regression algorithm, using the Scikit-Learn module in Python \citep{scikit-learn}. The adopted ML technique requires appropriate historical datasets for training. Logistic regression is one of simplest and widely-used ML algorithms for two-class classification. Logistic regression is a special case of linear regression where the target variable is dichotomous in nature. Dichotomous means that there are only two possible classes, e.g. yes/no or true/false. Logistic regression also predicts the probability of occurrence of a binary event utilising a logit function. 

Four training sets were constructed to enforce consistency in time and test robustness, each one corresponding to 6-, 12-, 18- and 24-hr forecast issuing time interval, because within a day the forecast reliability becomes more pronounced. The study takes as a reference the time of the largest flare event for each AR. For each issuing time interval, we consider the calculated $S_{l-f}$ or $G_{S}$ values of an AR before this reference time, as input data for the logistic regression. This framework allows us to quantify the prediction capabilities of the two morphological parameters.

Similarly to \cite{Korsos2016}, this study uses information on around 1000 ARs extracted from the Debrecen Sunspot Data Catalogue between 1996 and 2015 \citep{Baranyi2016}. The catalogue contains information including centroid position in various coordinate systems, area, and magnetic field of sunspots and sunspot groups. Derived from spacecraft observations, the catalogue has entries at each 1 hr for SDD\footnote{http://fenyi.solarobs.csfk.mta.hu/en/databases/SOHO/}, and 1.5 hr for HMIDD \footnote{http://fenyi.solarobs.csfk.mta.hu/en/databases/SDO/}. The GOES\footnote{https://www.ngdc.noaa.gov/stp/space-weather/solar-data/solar-features/solar-flares/x-rays/goes/xrs/} flare catalogue is used for information on the largest-intensity flare eruption of each AR.

For each issuing time interval, two thirds of the ARs were randomly extracted to create a training set. These ARs are labelled as true(1) and false(0) events, under two different binary classification definition models:
\begin{itemize}
\item {\bf1st model}: When the largest intensity flare of an AR is M- or X-class then this case is classified as true(1), otherwise B- or C-class flares are false(0).
\item {\bf 2nd model}: Based on the results of \cite{Korsos2016}, an event is true(1) if an AR is host to a M/X-class flare, satisfying 3$\leq$$S_{l-f}$, and 6.5$\leq$$\log(G_{S})$. Or, an event is true(1) if an AR was host to a B/C-class flare, satisfying $S_{l-f}$$>$3, and $\log(G_{S})$$<$6.5. Otherwise the cases are all labelled false(0).
\end{itemize} 
The two different classification models were chosen to study whether the two morphological parameters perform better, either with or without (2nd or 1st model) thresholds. Often, a well-chosen threshold adjustment(s) could improve prediction capabilities of a method, as a warning level or as a warning sign. Furthermore, in the case of both model approaches as described above, the set of $S_{l-f}$ and $G_{S}$ values associated with the remaining 1/3 ARs are not labelled and are provided as a test set only for the logistic regression algorithm training. In this manner, there is no overlap between training and testing. To ensure robustness of the results, we replicated 100 times the training and test datasets for 6/12/18 and 24-hr issuing time intervals, like e.g. \cite{Campi2019}.

\section{Analysis} \label{analyses}

\begin{figure}[t!]
\centering
\includegraphics[width=0.49\textwidth]{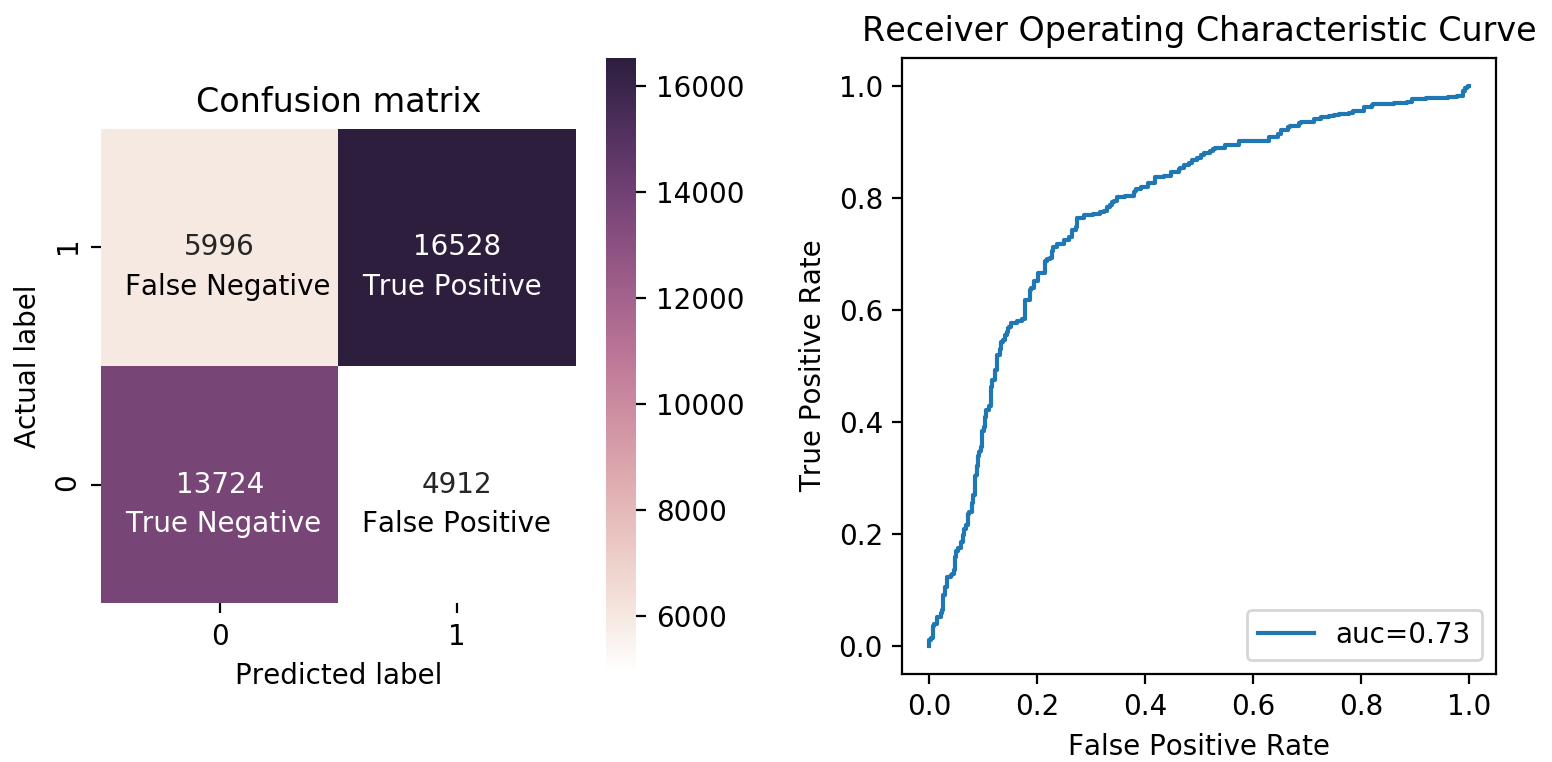}
\put(-240,110){(a)}
\includegraphics[width=0.49\textwidth]{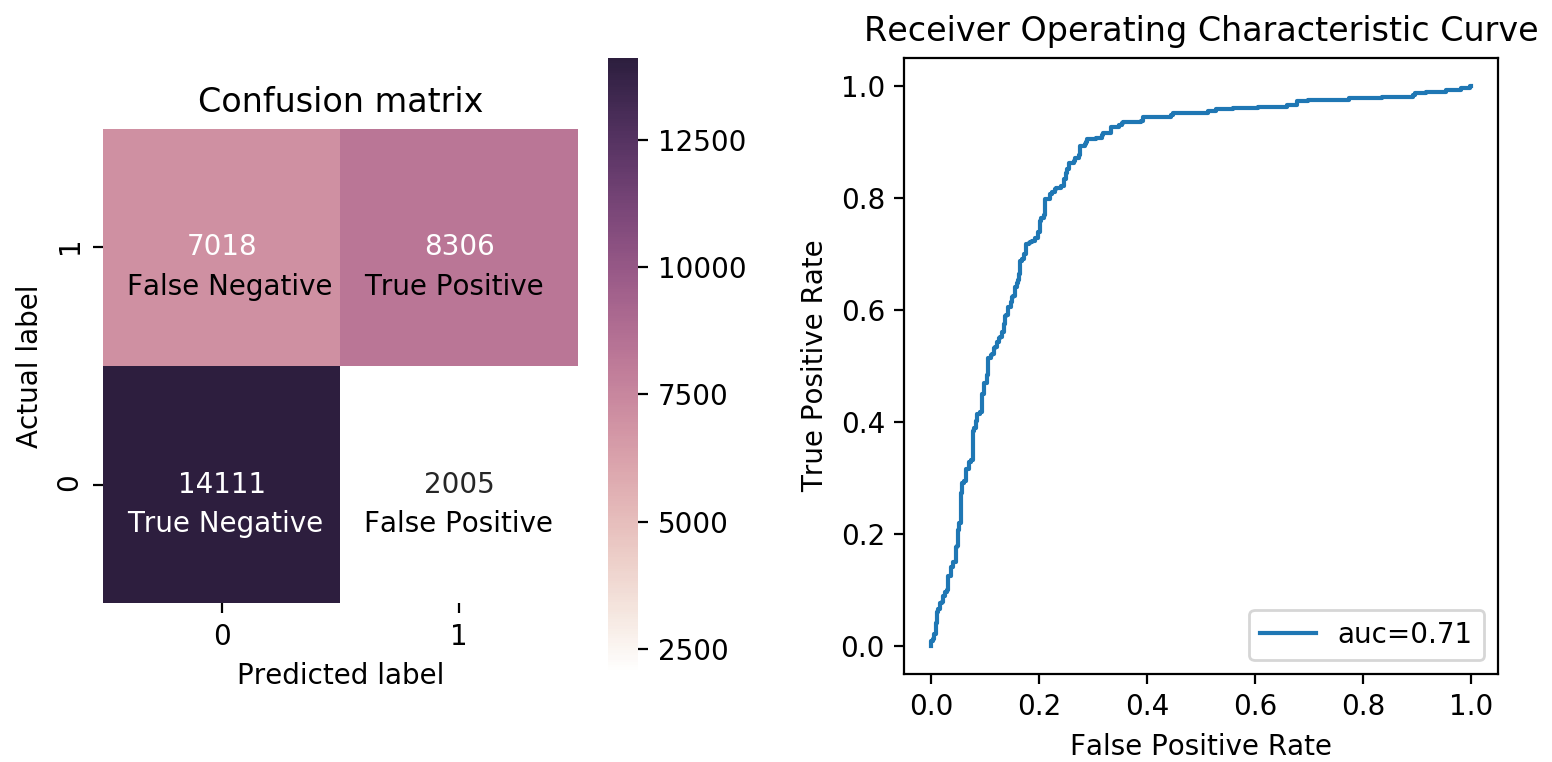}
\put(-240,110){(b)}\\
\includegraphics[width=0.49\textwidth]{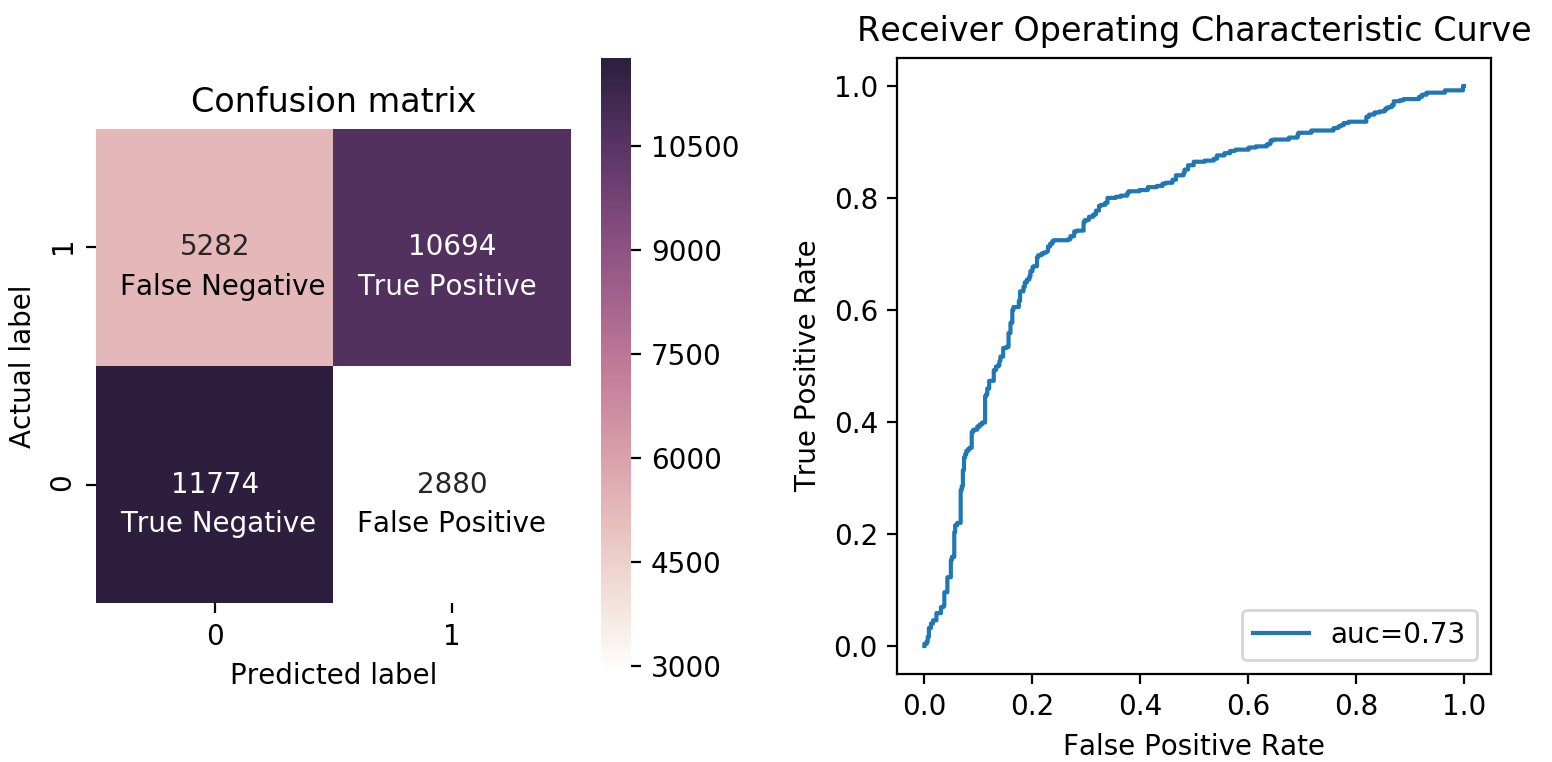}
\put(-240,110){(c)}
\includegraphics[width=0.49\textwidth]{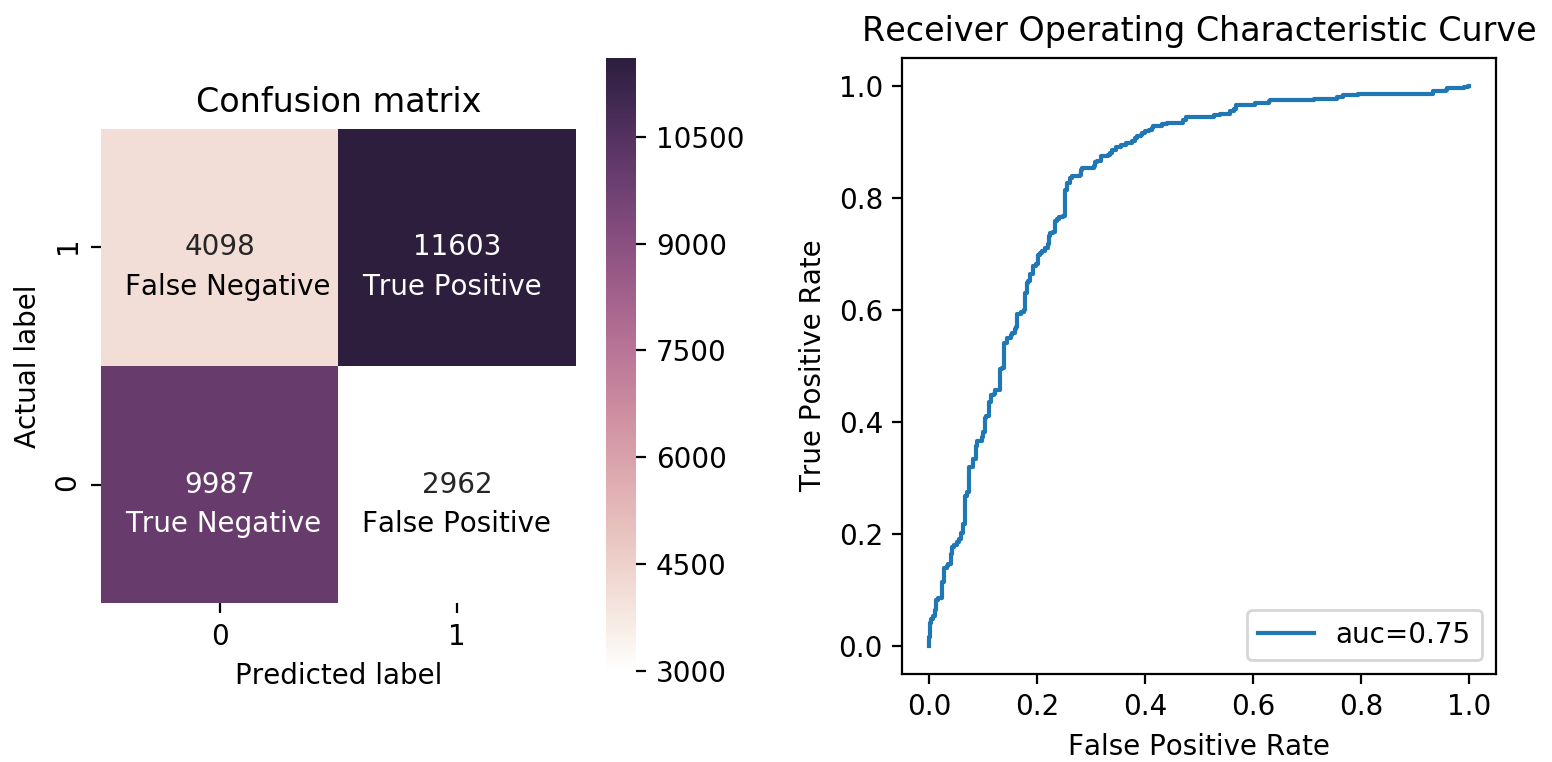}
\put(-240,110){(d)}
\caption{\label{Parameter}  Same as Fig.\ref{Flare}, but in the case of the 2$^{nd}$ model.}
\end{figure}

Solar flare prediction is affected by strong class imbalances, in that there are far more negative examples (labelled as N) than positive ones (labelled as P.) Therefore, we apply different metrics to measure the performance of the 1st and 2nd models. The performances of the two binary classifiers can be characterised by confusion matrixes in Figs.~\ref{Flare}-\ref{Parameter}.
Those confusion matrixes summarise the True Positive (TP), True Negative (TN), False Positive (FP), and False Negative (FN) predictions, we adopt different metrics
to quantify the impact performance of the $S_{l-f}$ and $G_{S}$ parameters in the case of both model approaches (1st and 2nd). The applied metrics are summarised in Table\ref{table1} for 6-, 12-, 18- and 24-hr forecast issuing times, and are:
\begin{itemize}
\item \underline{Accuracy} is the ratio of true positives plus true negatives over all events, or how often the TRUE prediction is correct: (TP+TN)/(P+N)
\item \underline{Recall}, also called the true positive rate or sensitivity, measures the proportion of actual positives that are correctly identified: TP/P
\item \underline{Specificity}, also called the true negative rate, measures the proportion of actual negatives that are correctly identified: TN/N
\item \underline{Precision}, also called positive predictive value. This is the ratio of true positives over all positive predictions: TP/(TP+FP). 
\item  \underline{Negative predictive value (NPV)} is the ratio of true negatives over all negative predictions: TN/(TN+FN). 
\item \underline{F1 score} is the harmonic mean between sensitivity (or recall) and precision (or ). It tells us how precise our two classifiers are, as well as how robust these are.
 A greater F1 score means that the performance of our model is better. Mathematically, F1 can be expressed as: 2$\cdot$(1/Recall+1/Precision)
\item \underline{True Skill Statistic (TSS)} is widely used to test the performance of forecasts \citep{McBride2000}. TSS will be the preferred performance metric when comparing results of the 1st and 2nd model approaches with different N/P ratios because this metric is independent from the imbalance ratio \citep{Woodcock1976,Bloomfield2012}. TSS takes into account both omission and commission errors. The TSS parameter is similar to Cohen's kappa approach \citep{Shao1995}, and compares the predictions against the result of random guesses. TSS ranges from -1 to +1, where +1 indicates perfect agreement. The zero or less value indicates that a performance no better than random \citep{Landis1977}.: TSS=TP/P-FP/N=Recall+Specificity-1
\end{itemize}

\begin{figure}
\centering
\includegraphics[width=0.6\textwidth]{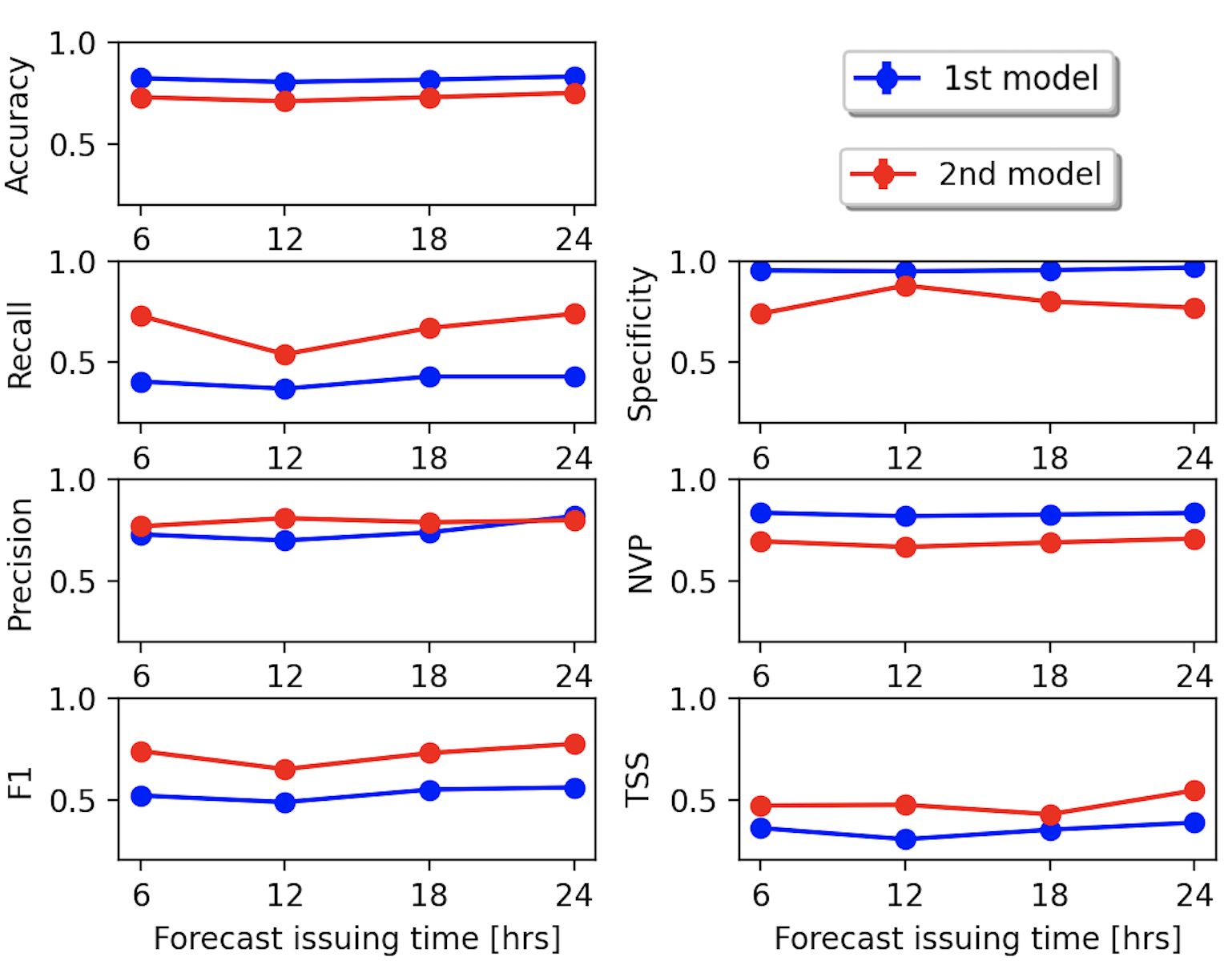}
\caption{\label{Metrics}  The evolution of selected metrics as a function of forecast issuing times for the 1st (blue) and 2nd (red) model.}
\end{figure}

These seven metric parameters are plotted as a function of forecast issuing times in Fig.~\ref{Metrics}, where the blue/red lines stand for the 1st/2nd model. Based on the values of Table\ref{table1} and Fig.~\ref{Metrics}, the two models have high accuracy for all forecast issuing times. In both models, the best accuracy is gained by the 24-hr prediction window. We emphasise that the accuracy is a meaningful measure only if the values of FP and FN would be similar in the confusion matrices of Figs.~\ref{Flare}-\ref{Parameter}. For dissimilar values, the other metrics must be considered in evaluating the prediction performance of the two models.

Next, we focus on the recall and specificity metrics, which show the probability whether a model captures the correct classification during all four intervals. The values of the specificity metric show that the two models are capable to correctly classify TN cases during all four intervals, especially in the case of the 1st model, which is greater than 90\%. Based on recall values, the TP classification of the 2nd model is 20\% more accurate than the 1st model for 6/12/18/24-hr forecast issuing times.

However, when the two models classify a new AR, then we do not know the true outcome until after an event. Therefore, we are likely to be more interested in the question what is the probability of a true decision of the two models.  This is measured by precision and NPV metrics. For the 1st model, the precision of the 24-hr prediction time is $\sim$10\% better than the other issuing time intervals, while the NPV values are $\sim$80\% in the case of four issuing time.
The precision and NPV values of the 2nd model are almost the same over all four prediction windows. Based on precision and NPV metrics, the 2nd model predict a TP event with higher probability than the 1st model, while the 1st model is better with the case of TN event. This is because the 2nd model discards some X- and M-class flares which do not satisfy the threshold conditions. Despite this, the 2nd model still could fairly predict a TN event with about 70\% probability.

The F1 and TSS metrics show that the 2nd model performs better than the 1st in the case of all of the prediction windows. This is an important aspect because the F1 and TSS are the most reliable scores in the presence of class imbalance. Intuitively, the F1 score is not as easy to understand as that of the accuracy, but it is usually more useful than accuracy, especially in our case, where we have an uneven class distribution. Namely, 77\% of the F1 score shows that the 24-hr flare prediction window is the most efficient in the case of the 2nd model approach. Furthermore, the above 0.4 values of TSS score of the 2nd model show that this method is a good prediction scheme, and, the defined accuracy values of the 2nd model can be considered as correct. 

 \begin{table*}
 \centering
\begin{tabular}{|c|cccc|cccc|}
\hline
{\bf Metrics}  & \multicolumn{4}{c|}{{\bf 1st model}}	& \multicolumn{4}{c|}{{\bf 2nd model} }	\\
\multirow{2}{*}{} &\multirow{2}{*}{6 hr} &\multirow{2}{*}{12 hr}&\multirow{2}{*}{18 hr}&\multirow{2}{*}{24 hr}	& 	 \multirow{2}{*}{6 hr} &\multirow{2}{*}{12 hr}&\multirow{2}{*}{18 hr}&\multirow{2}{*}{24 hr}	\\
&	 && &	& 	 &&&	\\\hline
Accuracy &0.82&0.81&0.82&  0.83	& 0.73	&0.71&0.73&0.75	\\\hline
Recall  &	0.41  &0.37&0.43&0.43	& 0.73	&0.54&0.67&0.74	\\\hline
 Specificity&0.95  &0.95&0.95&0.97	& 0.74	&0.87&0.80&0.77	\\\hline
 Precision&0.73&0.70&0.74& 0.82 	& 0.77	&0.81&0.79&0.80	\\\hline
  NPV&0.83&0.82&0.83& 0.84 	& 0.70	&0.67&0.69&0.71	\\\hline
    F1&0.52&0.48&0.52&0.56	  	& 0.75	&0.65&0.73&0.77	\\\hline
TSS&0.36&0.32&0.35&0.40	  	&0.47 	&0.42&0.47&0.51	\\\hline
\end{tabular}
 \caption{\label{table1}  
Flare prediction capabilities with six metrics in the case of the two model approaches, i.e. for 1st model and 2nd model.}
\end{table*}

We also use Receiver Operating Characteristic Curves (ROCs) to evaluated the results of the  binary logistic regression method for both models. In the ROC plots in Figures~\ref{Flare}-\ref{Parameter}, the sensitivity (the proportion of true positive results) is shown on the $y$-axis, ranging from 0 to 1 (0–100\%). The specificity (the proportion of false positive results) is plotted on the $x$-axis, also ranging from 0 to 1 (0–100\%). The area under the curve (AUC) is a measure of the test's performance at distinguishing positive and negative classes. In Figs.~\ref{Flare}-\ref{Parameter}, AUCs are above 0.7, or a capability to distinguish between positive class and negative class with more than 70\% likelihood over the 6-, 12-, 18- and 24-hr prediction time windows. From Fig.~\ref{Flare}, the 1st model shows similar AUC values during the four prediction windows. In the case of the 2nd model, the predicting probabilities are also similar based on the AUC values of Fig.~\ref{Parameter}. On further note that the predicting probabilities of the 2nd model are 10\% less than the 1st one, based on AUC values during the four prediction windows.

\section{Conclusion} \label{conclusion}
      
 \cite{Korsos2016} introduced the separation parameter $S_{l-f}$ and the sum of the horizontal magnetic gradient $G_{S}$ as potential indirect indicators of the measure of non-potentiality of the magnetic fields of solar active regions. They also proposed these two morphological parameters as potential new prediction proxy indicators complementing the traditional  Z\"urich, McIntosh or Mount Wilson classification schemes. 
       
In this work, a binary logistic regression machine learning approach is used to test and validate the flare prediction capability of the $G_{S}$ and $S_{l-f}$ morphological parameters. Two binary classification schemes are used. One scheme is based on a simple approach while implementing solely flare intensity, the second approach is a more sophisticated model based on both flare intensity and threshold values of the morphological parameters. This experimental approach is applied to a large set of $\sim$1000 ARs, with 100 repeats the datasets, over different forecast issuing time intervals of 6-, 12-, 18-, and 24-hr. Analysis of various performance metrics shows the following:

\begin{itemize}

\item The morphological parameters give more than 70\% flare prediction accuracy, based on logistical regression analysis. This result supports the findings of \cite{Kontogiannis2018} and \citep{Campi2019}, who conclude that the $G_{S}$ parameter has potential as an efficient predictor.
\item Based on the F1 scores and the True Skill Statistic metrics, the joint flare prediction efficiency of the $S_{l-f}$ and $G_{S}$ parameters is improved when the previously identified threshold values by \cite{Korsos2016} were also imposed. However, the 2nd model discards some X- and M-class flares which do not satisfy the threshold conditions. Despite of it, the 2nd model still could predict/classify an upcoming event with at least 70\% probability, based on the precision and NPV metrics. 
\item The best flare prediction capability of the two parameters is available with 24-hr forecast issuing time. This latter means that the $S_{l-f}$ and $G_{S}$ parameters with their thresholds are capable to predict an upcoming flare with 75\% accuracy a day before flare occurrence.
\item However, not just the 24 hrs prediction window has good metric scores, but also the ones with 6/12 and 18 hrs. This means that the $S_{l-f}$ and $G_{S}$ are together applicable for prediction purpose in a short- and long-term one.
\item The limitation of this study is that the applied data are extracted from a given sunspot database. Therefore, an other ML method (e.g. Convolutional Neural Network) that is trained on the same SDO/HMI intensity and magnetogram data, may assess further parameters to increase the predictive capability of the two morphological parameters.

\end{itemize}

We are aware that the two tested models are not perfect and so a natural question to ask is: how can we improve further them? In the future, we intend to further explore the application of these two warning parameters both from machine learning and physics perspectives: (a) fine tune the threshold conditions of 2nd model, (b) extend the application of the $S_{l-f}$ and $G_{S}$ parameters at different solar atmosphere heights, (c) train the employed machine learning model at different atmospheric heights for an even more accurate estimation of flare event time and flare event intensity, and (d) identify an optimal height range giving the earliest possible flare prediction, similar to the concept described by \cite{Korsos2020}.

\section*{Acknowledgements}
 
The authors are grateful to the Referees for constructive comments and recommendations which helped to improve the readability and quality of the paper. MBK and HM are grateful to the Science and Technology Facilities Council (STFC), (UK, Aberystwyth University, grant number ST/S000518/1), for the support received while carrying out this research. RE is grateful to STFC (UK, grant number ST/M000826/1) and EU H2020 (SOLARNET, grant number 158538). RE also acknowledges support from the Chinese Academy of Sciences President’s International Fellowship Initiative (PIFI, grant number 2019VMA0052) and The Royal Society (grant nr IE161153). JL acknowledges the support from STFC under grant No. ST/P000304/1.

\bibliographystyle{frontiersinSCNS_ENG_HUMS} 
\bibliography{test}

\end{document}